\documentclass[twocolumn,showpacs,prb,superscriptaddress]{revtex4}

\usepackage{graphicx}

\def\bea{\begin{eqnarray}}
\def\eea{\end{eqnarray}}
\def\a{\alpha}

\def\p{\partial} 
\def\nn{\nonumber}

\def\la{\langle}
\def\ra{\rangle}

\def\g{\gamma}

\def\f{\frac}

\def\A{\bf A}
\def\G{\bf G}

\begin{document}

\title{ Work distribution functions in polymer stretching experiments }
\author{Abhishek Dhar}
\email{dabhi@rri.res.in}
\affiliation{Raman Research Institute, Bangalore 560080}
\date{\today} 

\begin{abstract}
We compute the distribution of the work done in stretching a Gaussian
polymer, made of  $N$ monomers, at a finite rate. For a
one-dimensional polymer undergoing Rouse dynamics,  the
work distribution is a Gaussian and we explicitly compute the mean and width.
The two cases where the polymer is stretched, either  by constraining
it's end or by constraining the force on it, are examined. We 
discuss connections  to Jarzynski's equality and the fluctuation theorems.
\end{abstract}

\pacs{05.40.-a, 05.70.Ln}
\maketitle

\section{Introduction.}
\label{sec:intro}
Classical thermodynamics does not give us all the details
about a nonequilibrium process. For example consider a nonequilibrium
process during which we
perform work $W$ on a system kept in contact with a
heat bath at some fixed temperature $T$. The system starts from an
equilibrium state described by the temperature $T$ and some other
parameter, say $\lambda$ (e.g. volume). During the process the 
parameter changes from its initial value $\lambda_i$ to a final value
$\lambda_f$. At the end of the process the system need 
not be in equilibrium but will eventually relax to an equilibrium state
described by $T$ and $\lambda_f$.  
The second law then tells us that   
\bea
W \ge \Delta F
\label{ineq}
\eea
where $\Delta F$ is the difference of free energy between the two
equilibrium states. The equality holds if the process is
reversible. For an irreversible process what other 
information can one extract from a measurement of the
work done ?  First note that for specified initial and final values of
the parameter $\lambda$ and a fixed path $\lambda(t)$ connecting
them, the work done will not have a unique value. Every time we repeat
the process we will get a different work done because:  
(a) the initial microscopic   state we start from may be
different and (b) for a given initial microscopic state the time
evolution is not unique since the system is in contact with a heat
bath. Thus we will get a probability distribution for the
work done and it is of interest to examine the properties of
this distribution. 
Recently there has been a lot of interest on issues related to
properties of such distribution functions. 
Two very interesting results involving universal properties of these
distributions have been proposed. 

The first is a very surprising {\it exact equality} obtained by
Jarzynski \cite{jarz} which states that:
\bea
\la e^{-\beta  W} \ra = e^{-\beta \Delta F} 
\eea
where the average is over the work distribution function.
This result seems to hold very generally and should be compared with
the {\it{inequality}} in Eq.~\ref{ineq} that one obtains from usual  thermodynamics.

The second set of results are obtained when one looks at the probability
distribution of various  non-equilibrium quantities (including 
$W$) such as, for example, the entropy production. In this case some new fluctuation theorems
have been proposed \cite{evan1,evan2,galla}. These theorems were originally
derived for deterministic systems but have also been proved for
stochastic systems\cite{kurch,crook,dhar} These theorems look at the ratio of the
probabilities of positive to negative entropy
production during a nonequilibrium process and thus  give some measure
of ``second law violations'' which can be significant if one is
looking either at small systems or at small time intervals. There are two versions of the
fluctuation theorem, the steady state fluctuation theorem (SSFT) and the
transient fluctuation theorem (TFT). In the former case one looks at a
system in a nonequilibrium steady state and the average entropy
production rate is examined. In the transient fluctuation theorem a
system is initially prepared in thermal equilibrium and one looks at the
entropy produced in a finite time $\tau$.  
An important point to note is that the definition of entropy
production in small (non-thermodynamic) systems and in a
nonequilibrium situation  is somewhat ad-hoc and various definitions
have been used. A number of authors
\cite{wang,zon,zon2,menon,poggi,gold}have looked, both theoretically 
and in experiments, at fluctuations of quantities such as work, 
power flux, heat absorbed, etc. during a nonequilibrium process. 
In an interesting work it was shown by Crooks \cite{crook} that the Jarzynski
equality and the TFT are connected. 

Finding universal properties of various nonequilibrium distribution
functions is of obvious interest. At the same time  
the explicit forms of the distributions for different 
systems is clearly of interest too. Infact for systems such as polymers these are
experimentally accessible \cite{liph} and it seems plausible that they can give
information on the dynamics of the system. There has not been much
work in this direction. In a recent paper \cite{speck} Speck and Seifert have
shown that in the limit of slow driving the work distribution becomes
a Gaussian. They consider systems (which could be nonlinear) evolving
through Langevin dynamics. Apart from this work, most other explicit
calculations of nonequilibrium distribution functions  have considered
single particle systems. 

In this paper we consider the  well-known model of a flexible polymer
whose motion is governed by Rouse dynamics. We look at the work done when the polymer is
stretched at a finite rate. The work distribution
functions in different ensembles ( constant force and constant
extension) are computed explicitly and  the dependence of
the distributions on switching rates and system sizes is examined.  
We discuss our results in the context of Jarzynski's equality and the
fluctuation theorems (TFT).  
\section{Distribution of work in the constant extension ensemble.}
\label{sec:work}
We consider a one-dimensional Gaussian polymer whose energy is given
by:
\begin{equation}
H=\sum_{l=1,N+1} \f{k_l}{2} (y_l-y_{l-1})^2 
\label{Ham}
\end{equation}
with $y_0=0$ and $y_{N+1}=\alpha(t)$ which is a specified function of time. For
the moment we take the spring constants ${k_l}$ to be arbitrary. 
We assume the following Rouse dynamics for the chain:
\begin{equation}
\dot{y}_l=-\f{k_l}{\gamma} ( y_l-y_{l-1})+\f{k_{l+1}}{\gamma} (y_{l+1}-y_l)+\eta_l~~~l=1,2...N
\end{equation}
which can be written in matrix notation as
\begin{equation}
\f{dy}{dt}=-\f{1}{\gamma}{\A} y + \f{1}{\gamma} h(t) +\eta
\end{equation}
where $y^T=\{y_1,y_2....y_N\}$, $\eta^T=\{ \eta_1,\eta_2,...\eta_N \}$,
 $h^T=\{ 0,0,...k_{N+1} \alpha(t) \}$ and the noise satisfies
 $\la \eta_l(t) \ra=0$ and $\la \eta_l(t) \eta_m(t') \ra = 2/(\beta
 \gamma) \delta_{lm} \delta (t-t')$ . The matrix ${\A}$ is
tridiagonal with elements
 ${\A}_{l,l}=(k_l+k_{l+1}),~{\A}_{l,l+1}=-k_{l+1},~{\A}_{l,l-1}=-k_l$. 
The general solution of this equation is given by:
\begin{equation}
y(t)={\G}(t)y(0)+  \int_0^t dt' {\G}(t-t') \f{h(t')}{\gamma} +\int_0^t dt' {\G}(t-t') \eta(t') 
\label{solu}
\end{equation}
where ${\G}(t)=e^{-{\A} t/\gamma}$.
The work done, as defined by Jarzynski, is then given by:
\bea
W_J &=& \int_0^\tau \f{\p H}{\p \alpha} \dot{\alpha} dt \nn \\
&=& k_{N+1} \int_0^{\tau} (\alpha-y_N) \dot{\alpha} dt \nn \\
&=& \f{k_{N+1}}{2} (\alpha^2(\tau)-\alpha^2(0)) - \int_0^\tau
dt \dot{h}^T y.  \nn
\eea 
This work is  equal to the true mechanical work done by the
external force which we will call $W$ (thus in this case $W=W_J$). 
We now plug in the solution for $y_N$ from Eq.~\ref{solu} to get
\bea
&&W=\f{k_{N+1}}{2} (\alpha^2(\tau)-\alpha^2(0))-\int_0^\tau dt
\dot{h^T}[{\G}(t)y(0) \nn \\ &&+  \int_0^t dt' {\G}(t-t') \f{h(t')}{\gamma} +\int_0^t
dt' {\G}(t-t') \eta(t')]  
\label{Weq}
\eea
Since $W$ is linear in $y(0)$ and $\eta$ both of which are
Gaussian variables, it follows that the distribution of $W$ will also
be Gaussian. We then only need to find the mean and the second moment
which we now obtain.  We first state a few results on equilibrium
properties of the Gaussian chain. 
The equilibrium free energy of the chain is given by
\bea
Z(\alpha) &=& \int dy_1 dy_2...dy_N e^{-\beta H}~~{\rm{where}} \nn \\
H &=& \sum_{i=1}^{N+1} \f{k_i}{2} (y_i-y_{i-1})^2 \nn \\
&=& \f{1}{2} y^T {\A} y -h^T y +\f{1}{2}k_{N+1} \alpha^2 
\eea
This leads to the following equilibrium free-energy of the polymer (apart
from $\alpha$-independent constants).
\bea
F(\alpha)=\f{1}{\beta} ln(Z)&=& \f{1}{2} k_{N+1}\alpha^2 - 
\f{1}{2} h^T {\A}^{-1} h   \label{eqfr} \\
&=& \f{1}{2} \bar{k} \alpha^2 \nn
\eea
where $1/\bar{k}= 1/k_1+1/k_2+...1/k_{N+1}$. 
The mean positions of the particles and their fluctuations can be
easily computed at any force and are given by:
\bea
\la y \ra= {\A}^{-1} h \label{eqy} \\
\la \la (y - \la y \ra) (y^T - \la y^T \ra) \ra=\f{1}{\beta} {\A}^{-1} \label{eqysq}
\eea

{\bf{\underline{Mean and fluctuations of the work done}:}}

From Eq.~\ref{Weq} we get for the mean work done:
\bea
&&\la W \ra =\f{k_{N+1}}{2} (\alpha^2(\tau)- \alpha^2(0))-\int_0^\tau dt
\dot{h}^T[{\G}(t) \la y(0) \ra \nn \\ &&- \f{1}{\gamma} \int_0^\tau dt
 \int_0^t dt' \dot{h}^T {\G}(t-t') h(t')
\eea
We do  integration by parts so as to express everything in terms
of the rates $\dot{h}$. Using the equilibrium results in
Eq.~(\ref{eqfr},\ref{eqy}) we finally get:
\bea
\la W \ra = && F(\alpha(\tau))- F(\alpha(0)) \nn \\ &&+\int_0^\tau dt \int_0^t dt'
\dot{h}^T(t) {{\A}}^{-1} {\G}(t-t') \dot{h}(t')
\label{waveq}
\eea
The fluctuations of the work $\la (W-\la W \ra)^2 \ra =\sigma^2$ is given by:
\bea
\sigma^2 = \int_0^\tau dt \int_0^\tau dt' \dot{h}^T(t) {\G}(t)
\la [ y(0) - \la y(0) \ra ] [y(0)-\la y(0) \ra ]^T \ra &&\nn \\  \times {\G}(t')
\dot{h}(t')     +\int_0^\tau dt_1 \int_0^{t_1} dt_1'\int_0^\tau dt_2 \int_0^{t_2} dt_2'
\dot{h}^T(t_1) {\G}(t_1-t_1') && \nn \\  \times \la \eta(t_1')
\eta^T(t_2')\ra {\G}(t_2-t_2') \dot{h}(t_2)~~~~~~~~~~ && \nn
\eea
Using Eq.~(\ref{eqysq}) and the relation $\la \eta(t) \eta^T(t') \ra=
2/(\beta \gamma) \delta (t-t') {\bf I}$ this simplifies to:
\bea
\sigma^2 =\f{1}{\beta} \int_0^\tau dt \int_0^\tau dt' \dot{h}^T(t) {\G}(t)
{\A}^{-1}  {\G}(t') \dot{h}(t') +\f{4}{\beta \gamma} &&\nn \\  \times\int_0^\tau dt_1
\int_0^{t_1} dt_2 \int_0^{t_2} dt_2' 
\dot{h}^T(t_1) {\G}(t_1-t_2')  {\G}(t_2-t_2') \dot{h}(t_2) && \nn
\eea
Finally using the relation $\int_0^{t_2} dt_2' {\G}(t_1-t_2') {\G}(t_2-t_2') =
\f{\gamma {\A}^{-1}}{2} [ {\G}(t_1-t_2)-{\G}(t_1) {\G}(t_2) ]$ we get
\bea
\sigma^2=  \f{2}{\beta} \int_0^\tau dt \int_0^t dt'
\dot{h}^T(t) {\A}^{-1} {\G}(t-t') \dot{h}(t')
\label{sigmaeq}
\eea
The distribution of work done is then:
\bea
P(W)=\f{1}{(2 \pi \sigma^2)^{1/2}} e^{-\f{(W-\la W \ra)^2}{2 \sigma^2}}
\eea
As expected for a Gaussian process we find that 
\bea
\la W \ra - \Delta F= \beta \sigma^2 /2. 
\label{expect}
\eea
Note that in the present polymer model both the equilibrium free energy
and the average  work are independent of temperature while
the width of the distribution $\sigma^2$ depends linearly on temperature.
It is  easily verified that the work distribution satisfies the Jarzynski
equality $\la e^{-\beta W} \ra = e^{-\beta \Delta F}$. 
On the other hand, with the present definition of work, the
fluctuation theorem is not satisfied. However if 
we define the ``dissipated work''  $W_{diss}=W-\Delta F$ then the
distribution of $W_{diss}$, $\tilde{P}(W_{diss})$ satisfies the 
fluctuation theorem: 
\bea
\f{\tilde{P}(W_{diss})}{\tilde{P}(-W_{diss})}=e^{\beta W_{diss}}. \nn
\eea
\begin{figure}
\includegraphics[width=3in]{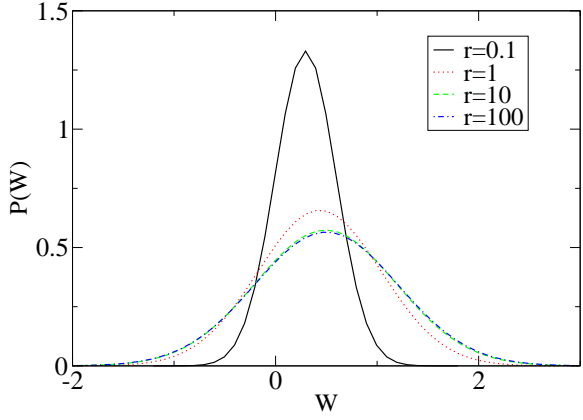}
\caption{Distributions of the work done in pulling a short polymer
  ($N=1$) at  different rates $r=\tau_R/\tau$.}
\label{n1} 
\end{figure}
Since $W_{diss}$ gives some measure of deviation from a quasistatic
and adiabatic process it seems reasonable to think of it as an entropy
production term which is what usually appears in the fluctuation
theorems. 

{\bf{\underline{Some special cases}:}}
Let us consider the case where all spring constants are equal $k_l=k$
and the let us assume that the polymer is pulled at a constant rate so that $\alpha (t) =
{a t}/{\tau}$. The effective spring constant is $k/(N+1)$ and so the
free energy of the polymer is $\f{k\a^2}{2(N+1)}$. From Eq.~\ref{sigmaeq}
we get the spread in the work done:
\bea
\sigma^2=\f{2  a^2 \gamma }{\beta \tau}  [k^2 {\A}^{-2}+
  \f{\gamma}{k \tau} k^3 {\A}^{-3} (e^{-{\A} \tau/\gamma}-1)]_{NN} 
\label{sigsp}
\eea 
where $[...]_{NN}$ denotes a matrix element. The average work can be
obtained from Eq.~\ref{expect}.  
In the two limits of very slow ($\tau \to \infty$) and very fast
($\tau \to 0$) processes, we get:
\bea
\sigma^2  &=& \f{2 k^2 a^2 \g {\A}^{-2}_{NN}}{\beta} \f{1}{\tau} \approx
\f{4 a^2 \gamma N}{\beta \pi^2 \tau} ~~~{\rm
  (Slow)}  \\
&=& \f{k a^2 N}{\beta (N+1)} ~~~~~~~~~~~{\rm (Fast)} 
\label{fast}
\eea
For an instantaneous pulling process the work done is simply
$W=k_{N+1} [ (\alpha(\tau)-y_N)^2-(\alpha(0)-y_N)^2]/2$ and the result
in  Eq.~\ref{fast} can be directly obtained.
\begin{figure}
\includegraphics[width=3in]{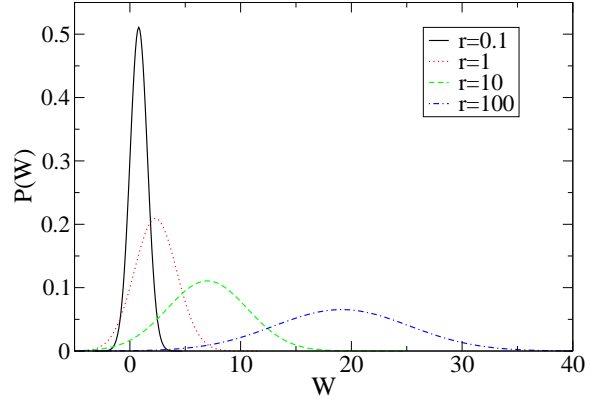}
\caption{Distributions of the work done in pulling a long polymer
  ($N=100$) at   different rates $r=\tau_R/\tau$.}
\label{n100} 
\end{figure}

It is instructive to plot the work distributions for different pulling
rates. To see the effect of the polymer length on the distribution we
consider two cases: (i) A short polymer with $N=1,~
k=1,~\gamma=0.1,~a=1,~\beta=1$ and  
(ii) a long polymer with $ N=100,~k=1,~ \gamma=0.1,~ a=10,~\beta=1$ (in arbitrary
units). In each case the parameter values are chosen so that the
change in equilibrium free energy given by   
$\Delta F= \f{k a^2}{2 (N+1)}$ is of the order of $1/\beta$ .
 The pulling rate has to be compared with
the relaxation time of the polymer which is  given by
$\tau_R=\gamma/\lambda_{sm}$ where $\lambda_{sm}=4k\sin^2(\f{\pi}{2(N+1)})$ is the smallest
eigenvalue of ${\A}$. For large $N$ we get $\tau_R=\gamma N^2/(k
\pi^2)$. For the cases (i) and (ii) we numerically evaluate Eq.~\ref{sigsp} for pulling rates
$r=\tau_R/\tau=0.1,~1,~ 10,~100$. The resulting distributions are
plotted in Fig.~\ref{n1} and Fig.~\ref{n100}.
For the long polymer the probability of negative work realizations is
quite small. We can increase their probability by increasing the
temperature which broadens the distributions while keeping the mean
unchanged. 
\section{Distribution of work in the constant-force ensemble.}
Next we compute the work distribution in the constant-force ensemble.  
Instead of constraining the end of the polymer we apply
a  time-dependent force $f(t)$ on it. The time-dependent
Hamiltonian of the system is now given by:
\bea
H=\sum_{l=1,N} \f{k_l}{2} (y_l-y_{l-1})^2 -f(t)y_N
\eea
and the equations of motion are again:
\bea
\f{dy}{dt}=-\f{1}{\gamma}{\A}
 y + \f{1}{\gamma} h(t) +\eta \nn 
\eea
with $h^T(t)=[0,0,...f(t)]$. In this case we note that the the
generalized work $W_J$, as defined by Jarzynski, is given by
$W_J=-\int_0^\tau dt y_N \dot{f}(t) dt$ and is not equal to the true
mechanical work done on the system 
which is $W=\int_0^\tau dt f(t) \dot{y}_N$. In this paper we will compute the
distribution of the true mechanical work $W$. We again find that $W$ has a Gaussian
distribution with the following mean and variance:
\bea
\la W \ra = \f{1}{\gamma} \int_0^\tau dt h^T(t) h(t)-\f{1}{\gamma^2}
\int_0^\tau dt \int_0^{t} d t' h^T(t) ~~~~&&\nn \\ \times {\G}(t-t') {\A} h(t') -\f{1}{g}
\int_0^\tau dt h^T(t) {\G}(t) h(0) &&\nn \\
\sigma^2 = \f{2}{\beta \gamma} \int_0^\tau dt h^T(t) h(t)-\f{2}{\beta \gamma^2}
\int_0^\tau dt \int_0^{t} d t' h^T(t) && \nn \\ \times {\G}(t-t') {\A} h(t') \nn
\eea
For $f(0)=0$, we get $\la W \ra = \f{\beta} {2} \sigma^2$ which means
that $P(W)$ satisfies the fluctuation theorem. It is then natural to
again ask if $W$ is some measure of entropy production. If this was so
then $W$ should vanish for an adiabatic process. To check this we
first express $\la W  \ra$ in terms of the rate $\dot{h}(t)$. We get
\bea
 \la W \ra   = \f{1}{2} [ h^T(\tau) {\A}^{-1} h(\tau)-h^T(0) {\A}^{-1}
  h(0) ] && \nn \\
-h^T(\tau)  \int_0^\tau dt {\A}^{-1} {\G}(\tau - t) \dot{h}(t)   
+ \int_0^{\tau}dt \int_0^t dt' \dot{h}^T(t) && \nn \\ \times {\A}^{-1}  {\G}(t-t')\dot{h}(t')~~~~~&& \nn
\eea
Hence for an adiabatic process we get $\la W \ra= -\Delta {\G}(f)$
where ${\G}(f) = -h^T{\A}^{-1}h/2$ is the polymer free energy in the
constant force ensemble.
Thus in this case
$\la W \ra$ {\it is not zero} for an adiabatic process and so it is
not an obvious measure of entropy production {\it even though it does satisfy
the fluctuation theorem}. 

Interestingly, since $\sigma^2 = 2 \la W
\ra/ \beta $, thus even for an adiabatic process, the
work-distribution {\it does not tend} to a $\delta$ function as one might
naively expect. However in the thermodynamic limit $N \to
\infty$ the width approaches zero as $\sigma \sim 1/N^{1/2}$ so in
this limit the usual expectation is indeed satisfied.     

\section{Conclusions.}
In conclusion, in this paper we have computed explicitly the
distribution  of work done when a  polymer is stretched at a
finite rate. We examine different ensembles and look at different
definitions of work. As has been noted by earlier authors, for different
definitions of the work, the corresponding distributions can have
quite different properties. In the constant extension ensemble the
generalized work $W_J$ is the same as the true mechanical work $W$ and
Jarzynski's identity is satisfied. In this case, the fluctuation
theorem is satisfied by a different quantity $W_{diss}$ which does seem
like a quantity which gives some measure of entropy production.    
In the constant force ensemble, $W_J$ is different from the true work
$W$ which  does not satisfy the Jarzynski identity.  
On the other hand the distribution of $W$ satisfies a fluctuation-theorem like
relation. However in this case we find that it is not possible to identify the
work as a measure of entropy production. 

As a practical use, the Jarzynski identity has been proposed as an
efficient method for computing equilibrium free energy profiles from
nonequilibrium measurements, both in simulations and experiments \cite{hummer,felix}.   
For the specicific case of polymers, we expect that a combination of
simulations and our exact results on the work distribution, should lead to
better estimates on efficiency and of errors \cite{zuckerman,gore,fox} (and how they depend on
rates and system sizes) involved while using the nonequilibrium
methods in free energy computations.

For non-Gaussian models of polymers such as, for example, semiflexible
polymers, the work distribution function is likely to be non-Gaussian. For
small pulling rates one again expects a Gaussian distribution. It will be
interesting to compute such distributions explicitly and  study their
dependences on rates, system sizes and other parameters such as the
rigidity of the polymer.


\begin{thebibliography}{**}
\bibitem{jarz} C. Jarzynski, \emph{Nonequilibrium equality for free energy
  differences}, Phys. Rev. Lett. {\bf 78}, 2690 (1997); 
  C. Jarzynski, \emph{Equilibrium free-energy differences from
  nonequilibrium measurements: A master-equation approach}, Phys. Rev. E {\bf 56}, 5018 (1997).   
\bibitem{evan1}  D. J. Evans, E G. D. Cohen, and G. P. Morriss, \emph{
  Probability of second law violations in shearing steady states}, 
  Phys. Rev. Lett. {\bf 71}, 2401 (1993).
\bibitem{evan2} D. J. Evans and D. J. Searles, \emph{ Equilibrium
  microstates which generate second law violating steady states}, 
  Phys. Rev. E {\bf 50}, 1645 (1994). 
\bibitem{galla} G. Gallavotti and E. G. D. Cohen, \emph{Dynamical ensembles in nonequilibrium statistical mechanics},   Phys. Rev. Lett. {\bf 74}, 2694 (1995).
\bibitem{kurch} J. Kurchan, \emph{Fluctuation theorem for stochastic
  dynamics}, J. Phys. A: Math. Gen. {\bf 31}, 3719 (1998); 
J. L. Lebowitz and H. Spohn, \emph{A Gallavotti-Cohen-type symmetry in the
  large deviation functional for stochastic dynamics},
  J. Stat. Phys. {\bf 95}, 333 (1999).  
\bibitem{crook} G. E. Crooks, \emph{Entropy production fluctuation theorem
  and the nonequilibrium work relation for free energy differences},
  Phys. Rev. E {\bf 60}, 2721 (1999).  
\bibitem{dhar} O. Narayan and A. Dhar, \emph{Reexamination of experimental
  tests of the fluctuation theorem}, J. Phys. A: Math. Gen. {\bf 37} 63 (2004).
\bibitem{wang} G. M. Wang et al., \emph{Experimental demonstration of
  violations of the second law of thermodynamics for small systems and
  short time scales}, Phys. Rev. Lett. {\bf 89}, 050601 (2002).
\bibitem{zon} R. van Zon and E. G. D. Cohen, \emph{Extension of the
  Fluctuation Theorem }, Phys. Rev. Lett. {\bf   91}, 110601 (2003).
\bibitem{zon2} R. van Zon, S. Ciliberto and E. G. D. Cohen, \emph{ Power
  and Heat Fluctuation Theorems for Electric Circuits},  
  Phys. Rev. Lett. {\bf 92}, 130601 (2004)  
\bibitem{menon} K. Feitosa and N. Menon, \emph{Fluidized Granular Medium as an Instance of the Fluctuation Theorem}, Phys. Rev. Lett. {\bf 92}, 164301 (2004).  
\bibitem{poggi} S. Aumaître, S. Fauve, S. McNamara and P. Poggi,
 \emph{ Power injected in dissipative systems and the fluctuation theorem}, Eur. Phys. J. B
{\bf 19}, 449 (2001).
\bibitem{gold} W. I. Goldburg, Y. Y. Goldschmidt and H. Kellay,
  \emph{Fluctuation and Dissipation in Liquid-Crystal Electroconvection },  
  Phys. Rev. Lett. {\bf 87}, 245502 (2001).  
\bibitem{liph} J. Liphardt, S. Dumont, S. Smith, I. Tinoco, and
C. Bustamante, \emph{Equilibrium information from nonequilibrium
measurements in an experimental test of Jarzynski's equality}, Science {\bf 296}, 1832 (2002). 
\bibitem{speck} T. Speck and U. Seifert, \emph{Distribution of work in
  isothermal nonequilibrium processes}, arXiv:cond-mat/0406279/, (2004). 
\bibitem{hummer} G. Hummer and A. Szabo, \emph{Free energy reconstruction
  from nonequilibrium single-molecule pulling experiments}, Proc. Natl. Acad. Sci {\bf
  98}, 3658 (2001).  
\bibitem{felix} F. Ritort, \emph{Work fluctuations, transient violations of
  the second law and free-energy recovery methods: perspectives in
  theory and experiments}, Poincar´e Seminar {\bf 2}, 195 (2003)
\bibitem{zuckerman} D. M. Zuckerman and T. B. Woolf, \emph{ Theory of a
  systematic computational error in free energy differences}, 
  Phys. Rev. Lett. {\bf 89}, 180602 (2002).
\bibitem{gore} J. Gore, F. Ritort and C. Bustamante, \emph{Bias and error
  in estimates of equilibrium free-energy differences from
  nonequilibrium measurements}, Proc. Natl. Acad. Sci. {\bf 100}, 12564 (2003).
\bibitem{fox} R. F. Fox,  \emph{Using nonequilibrium measurements to
  determine macromolecule free-energy differences}, Proc. Natl. Acad. Sci. {\bf 100}, 12537 (2003).
\end{thebibliography}

\end{document}